\documentclass[12pt]{article}
\usepackage{graphicx}
\usepackage{amsmath,amssymb}
\usepackage[latin2]{inputenc}

  \def\be{\begin{equation}}
\def\ee{\end{equation}}  
\def\ba{\begin{array}{c}}

\def\ea{\end{array}} 
\def\bea{\begin{eqnarray}}
\def\eea{\end{eqnarray}}
 
\begin{document}

%%%%%%%%%%%%%% FOR EDITORIAL USE ONLY!!! %%%%%%%%%%%%%%%
\begin{center}
{\Large \bf An explicitly solvable model of the spontaneous
$PT$-symmetry breaking\\} $$ $$ {\small V\'i{}t
Jakubsk\'y\footnote{jakub@ujf.cas.cz},
 Miloslav Znojil\footnote{znojil@ujf.cas.cz}}
\end{center}

\begin{abstract}
We contemplate the pair of the purely imaginary delta-function
potentials on a finite interval with Dirichlet boundary
conditions. The two parameter model exhibits nicely the expected
quantitative features of the unavoided level crossing and of a
"phase-transition" complexification of the energies. Combining
analytic and numerical techniques we investigate strength- and
position-dependence of its spectrum.
\end{abstract}

   %\section*{Introduction}

\section{The Model}
Systems with point interactions play an important role in quantum
mechanics since they keep their exact solvability while describing
real systems~\cite{albeverio1}. Conception of point interactions
in pseudo-Hermitian quantum mechanics appeared in various
works~\cite{albeverio2},~\cite{fei},~\cite{cervero},~\cite{znojil1}.

The aim of present article is to describe spectral properties of
one particle confined in infinite square well with two point
interactions of imaginary strength. The corresponding time-less
Schr\"odinger equation reads
 \be H\psi=\left(-\triangle -i\xi\delta(x+a)+i\xi\delta(x-a)\right)\psi=E\psi.\label{1}\ee
where $H$ acts on continuous functions from $L_2((-1,1),dx)$
vanishing in $\pm1$ and whose derivatives have an imaginary step
$\psi'(\pm a_{R})-\psi'(\pm a_{L})=\pm i\xi\psi(\pm
a)$\footnote{Indexes $L,\ R$ denote direction in which we approach
the point.}. Considering energy $E$ as a general complex number,
following the standard procedure we arrive at the secular equation
 \be \sinh 2u+\frac{\xi^2}{4u^2}\left(\sinh(4ua-2u)+\sinh2u-2\sinh 2ua\right)=0,\label{2}\ee
where $u$ is defined as $E=-u^2$. Invariance of the equation with
respect to complex conjugation of its root was rather expected due
to the $P$-pseudo Hermiticity of the hamiltonian
($H^{\dagger}=PHP$). Due to the lack of physically relevant
interpretation for non-real eigenvalues $E$, we will be interested
 in real energies only. In that case, (\ref{2}) acquires the
following simplified form,

 \be\sin 2k+\frac{\xi^2}{k^2}\sin 2ka\sin^2 k(1-a)=0,\ \ k=Im(u).\label{3}\ee
Neither (\ref{2}) nor (\ref{3}) can be solved without numerical
methods. Before presentation of numerical solution, we will focus
on qualitative properties of (\ref{3}) that can be derived
analytically.

 \section{Spectral lines - robust vs. fragile}
Performing a formal manipulation, the equation (\ref{3}) can be
rewritten,
 \be\xi^2=-\frac{\sin2k}{\sin 2ka}\frac{k^2}{\sin^2k(1-a)}.\label{5}\ee
In (\ref{1}), we required $\xi$ to be non-negative real number. It
restricts the domain of $\xi=\xi(k)$ since the right-hand side of
(\ref{5}) must be positive.

Let us take the following set
 \be{\cal{J}}=\left(J_a^+\bigcap I^-\right)\bigcup\left(J_a^-\bigcap I^+\right)\bigcup M,\label{6}\ee
where
 $$ k\in I^{\pm}\Leftrightarrow \pm \sin2k>0,\ \ k\in J_a^{\pm}\Leftrightarrow \pm \sin2ka>0,
 \ k\in M\Leftrightarrow \sin2k=\sin2ka=0  .\nonumber$$
Assuming that $k\in {\cal{J}}$, we define $\xi$ to be equal to its
limit in the points where $\sin k(1-a)=0$. Then the function
$\xi=\xi(k)$ is continuous on the open interval $J_a^{\pm}\bigcap
I^{\mp}$. If $k\in M$, equation (\ref{3}) is fulfilled for any
real $\xi$.

Let us discuss now the behavior of $\xi$ at the boundaries of the
subintervals of ${\cal J}$. The numerator $\sin 2k$ equals zero in
one boundary point at least. It is due to $a<1$ and $\xi=0$ in
this case. We will say that the boundary point is of the first
type. On the other hand when $\sin 2ka$ vanishes, $\xi$ tends to
infinity and the boundary is of the second type.

Thus, we can observe three kinds of spectral lines. The first one
is supported by an interval from ${\cal{J}}$ with both boundaries
of the first type. It reaches its maximum $\xi_{max}$ on the
interval so it is bounded. The spectral lines of the second type
tend to infinity and are supported by subintervals of
$J_a^{\pm}\bigcap I^{\mp}$ with mixed types of boundaries. The
third kind is unbounded as well and is supported by $k\in M$. That
are semi-lines perpendicular to $k$-axes. In terminology used in
\cite{znojil2}, the energy levels of the first type are called
{\bf fragile} while the remaining ones are {\bf robust} (see Fig.
1).

\newsavebox{\figa}
    \savebox{\figa}{
    \rotatebox{0}{\scalebox{0.8}{
    \includegraphics*{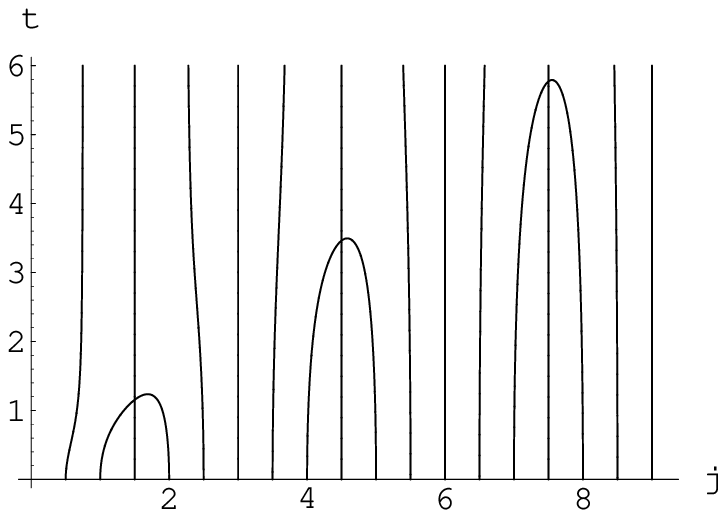}
    }}
    }
\newsavebox{\figb}
    \savebox{\figb}{
    \rotatebox{0}{\scalebox{0.8}{
    \includegraphics*{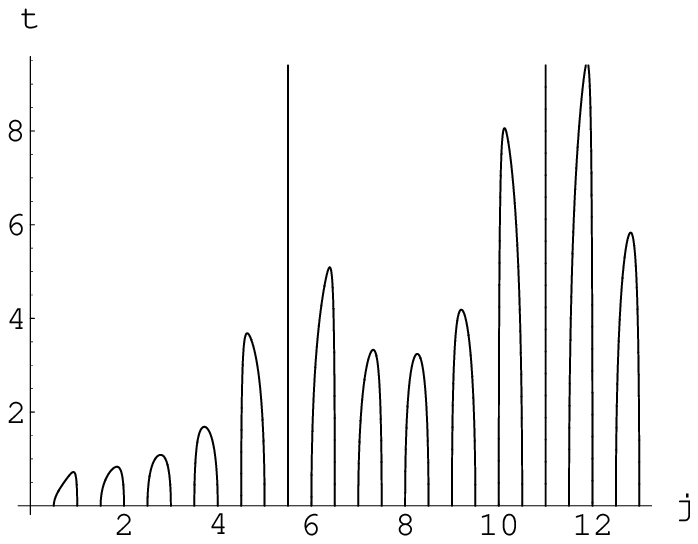}
    }}
    }
\newsavebox{\figc}
    \savebox{\figc}{
    \rotatebox{0}{\scalebox{0.6}{
    \includegraphics*[59mm,102mm][151mm,193mm]{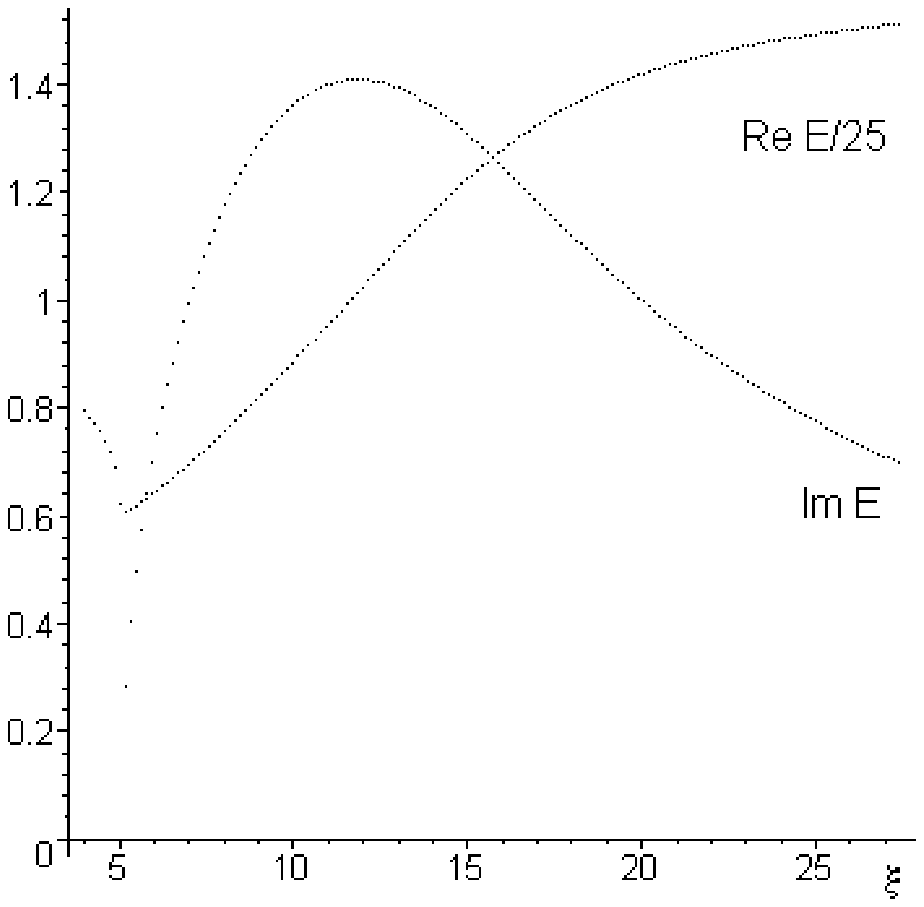}
    }}
    }
 \begin{figure}[t]
\begin{center}
 %\framebox[20em]{\rule{0pt}{50mm}}
 \usebox{\figa}
 \end{center}
 \vspace{-2mm} \caption{The function $\xi=\xi(k)$ for $a=2/3$. All three types of spectral lines are present, $\xi=5t$, $k=j\pi$. }
 \end{figure}

The third class of spectral lines corresponds to wave functions
having nodes at the interaction points and it appears only for
rational $a$. On two intervals $j_{\pm}\subset J_a^{\pm}\bigcap
I^{\mp}$ that are connected by $k\in M$, spectral lines of the
first and the third kind are stuck together. We speak about {\bf
unavoided level crossing} in this case. Due to simple form of
(\ref{5}), the level crossing can be described quantitatively. Let
us take $a=\frac{P}{Q} $ where $P$ and $Q$ are coprime integers.
The level crossing appears if $2k=l Q\pi$ and $ l (Q-P)$ is odd
number, $l$ is integer. In that case, the crossing of spectral
lines occurs for
 $$\xi_{cross}=\frac{l \pi Q^{\frac{3}{2}}}{2 P^{\frac{1}{2}}\sin \frac{l\pi(Q-P)}{2}}. $$

The above analysis  provided us a qualitative knowledge about
behavior of $k=k(\xi)$ implicitly given by (\ref{2}). The fragile
spectral lines cease to be real when $\xi$ exceeds a critical
value $\xi_{max}$. Thus for any $a$, we can find a sequence of the
critical couplings ${\xi_{crit,n}}$ where $n$ is the excitation of
the higher energy from the couple of merging energy levels. The
sequences of critical couplings have their minimum $\xi_{min}\sim
2.4931$ for $a\sim0.3335$. It is an interesting feature of our
model that the sequence is not monotonic. We call it {\bf an
overstepping complexification} of energy levels (see Fig 2.). The
numerical solution $k=k(\xi)$ of (\ref{2}) in the supercritical
regime is in Fig.3.
\begin{figure}[t]
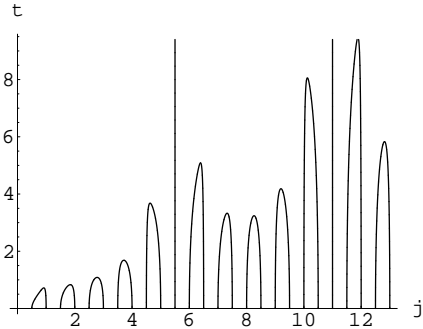

\begin{center}
%\framebox[20em]{\rule{0pt}{50mm}}
\usebox{\figb}
\end{center}
\vspace{-2mm} \caption{Overstepping complexification of energy
levels for $a=\frac{1}{11}$, $\xi=8t$, $k=j\pi$.}
\end{figure}

\begin{figure}[t]
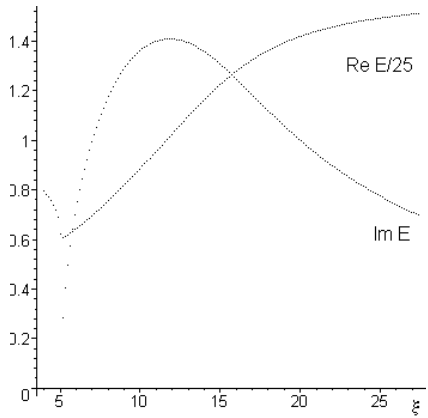

\begin{center}
%\framebox[20em]{\rule{0pt}{50mm} }
\usebox{\figc}
\end{center}
\vspace{-2mm} \caption{Real and imaginary part of $E=-u^2(\xi)$
after crossing exceptional point for $a=\frac{1}{2},\ u(0)=\pi$.}
\end{figure}

\section{Conclusion and Outlook}
The presented model of a particle captured in an infinite square
well with two point $PT$-symmetric interactions~(\ref{1}) offers
an interesting spectral behavior. We studied a parametric
dependence of energy levels on a position and strength of the
point impurities. The analysis provided us a good qualitative
control over various phenomena appearing in the spectrum. We
described presence of fragile and robust spectral lines as well as
appearance of unavoided level crossing of energy levels. These
analytical observations were supported by numerical results. In
the numerically obtained data, we observed a new phenomenon of
"overstepping" complexification where merging of energy levels is
not sequent with respect to the excitation but proceeds at random.

The article is a continuation of research started in
\cite{znojil1}. In this sense, it is straightforward to find
perturbation series for energy in the case of generic $a$. It has
been shown that for $a=1/2$, the effect of non-Hermitian
interaction was negligible for high excitations. This fact enables
approximate separability of positive-definite metric performed
originally in \cite{mostafazadeh} and consequent computation of
physically relevant quantities like position expectation value and
its time development.

\bigskip
{\small Work supported from the budget of the AS CR project AV 0Z
1048901. Participation of M. Z. partially supported by grant GA AS
{C}R, grant Nr. 104 8302 . Participation of V. J. supported by
grant MSMT FRVS {C}R, grant Nr.2388G-6}

\end{document}